\documentstyle[12pt]{article}

\begin{document}

\author{{\bf R. Mignani}$^{\ast }${\bf and R.Scipioni}$^{\circ }$ \\
$^{(\ast )}$Dipartimento di Fisica ''E. Amaldi''\\
Universita' di Roma ''Roma Tre'' \\
Via della Vasca Navale, 84 \\
00146 Roma, Italy\thanks{%
mignani@fis.uniroma3.it}\\
$^{(\circ )}$Department of Physics and Astronomy\\
The University of British Columbia \\
6224 Agricultural Road\\
Vancouver, B.C., Canada V6T 1Z1\thanks{%
scipioni@physics.ubc.ca}}
\title{{\bf On the solutions of the Cartan equation in Metric Affine Gravity}}
\maketitle

\begin{abstract}
In the Tucker-Wang approach to Metric Affine gravity we review some
particular solutions of the Cartan equation for the non-riemannian part of
the connection. As application we show how a quite general non Riemannian
model gives a Proca type equation for the trace of the nonmetricity 1-forms $%
Q$.\newline
\end{abstract}

\newpage

\section{INTRODUCTION}

Einstein's theory of gravity which was developed more than eighty years ago
provides an elegant and powerful formulation of gravitation in terms of a
pseudo-Riemannian geometry. In the variational approach, Einstein's
equations are obtained by considering variations, with respect to the
metric, of the Hilbert-Einstein action, i.e. the integral of the curvature
scalar, associated with the Levi-Civita connection, on the spacetime-volume
form . Einstein assumed that the connection was metric compatible and
torsion free; a position which is natural but not always convenient. In
particular, a number of recent developments in physics suggest the
possibility that the treatment of spacetime might involve more than just a
Riemannian structure. Let us quote some of them:\newline
\newline
1] The vain effort to quantise gravity, which is perhaps so far the
strongest piece of evidence for going beyond a geometry dominated by the
classical concept of distance.\newline
\newline
2] The generalisation of the theory of elastic continua with structure to
4-dimensional spacetime, that provides physical interpretations of the
non-Riemannian structures which emerge in the theory [1,2].\newline
\newline
3] The description of hadronic (or nuclear) matter in terms of extended
structures [3,4].\newline
\newline
4] The study of the early universe, in particular singularity theorems, the
problem of the unification of interactions and the related problem of
compactification of dimensions, and models of inflation with dilaton-induced
Weyl covector [5].\newline
\newline
Moreover, at the level of the so called string theories there are hints
[6-9] that by using non-Riemannian geometry we may accommodate the several
degrees of freedom coming from the low-energy limit of string interactions
in terms of a non \ metric-compatible connection with torsion. It is
interesting to observe that, since string theories are expected to produce
effects which are at least in principle testable at low energies, there may
be chances to obtain non-Riemannian models with predictions which can
somehow be tested; some models may have even some effects on astronomical
scales [10-20]. For instance, recently models have been proposed that permit
to account for the so-called dark matter by invoking non-Riemannian
gravitational interactions [21]. There are several approaches to
non-Riemannian gravity: perhaps one of the most popular is that which uses
gauge field theories [22-25].\newline
Soon after Einstein proposed his gravitational theory, Weyl found an
extension to it, able to include electromagnetism in a unified way [26].
Weyl's theoretical concept was the so called {\em gauge invariance of length}%
. To that purpose, Weyl extended the geometry of spacetime from the
Levi-Civita connection to a new space (''{\it Weyl space''}) with an
additional covector $Q=Q_{a}e^{a}$ , where $e^{a}$ denotes the field of
coframes of the four-dimensional manifold.\newline
The Weyl connection 1-form reads: 
\[
\Gamma ^{W}{}_{\alpha \beta }=\Gamma _{\alpha \beta }+\frac{1}{2}(g_{\alpha
\beta }Q-e_{\alpha }Q_{\beta }+e_{\beta }Q_{\alpha }) 
\]
The Weyl form is related to the so-called {\it non-metricity of the spacetime%
}. If we write the interval in the form: 
\[
ds^{2}=g_{\mu \nu }dx^{\mu }dx^{\nu } 
\]
the square length of a generic vector $V$ can be written as: 
\[
V^{2}=g_{\mu \nu }V^{\mu }V^{\nu }=g(V,V) 
\]
where $g$ is the (2,0) symmetric tensor defined by: 
\[
g=g_{\mu \nu }\,e^{\mu }\otimes e^{\nu } 
\]
We find that the covariant derivative of $V^{2}$ with respect to a generic
vector $X$ gives: 
\[
\nabla _{X}V^{2}=(\nabla _{X}g)(V,V)=Q_{ab}(X)V^{a}V^{b} 
\]
where $Q_{ab}(X)=({\nabla _{X}}g)(X_{a},X_{b})$, $Q^{a}{}_{a}=Q$ and we
assumed that ${\nabla _{X}}V=0$.\newline
In a spacetime with nonmetricity, the length of a vector changes if we
parallely transport the vector along a curve whose tangent vector is $X$.%
\newline
In Weyl's theory the field $Q$ is identified with the electromagnetic
potential $A$.\newline
Subsequently it was found that Weyl's theory is not viable. However the
concept of gauge invariance survived. In particular the concept of local
gauge invariance flourished in the field of theoretical particle physics.
Consider a particle, described, in quantum mechanics, by a wave function $%
\Psi $. Then, as is well known, postulating the local invariance of the
theory under the $U(1)$ abelian group, namely 
\[
\Psi \rightarrow e^{i\alpha (x)}\Psi , 
\]
(with $\alpha (x)$ a function of spacetime coordinates), permits to get the
electromagnetic interaction, and therefore to construct the whole classical
Dirac-Maxwell theory for a charged particle in an electromagnetic field.

In 1954 Yang and Mills generalised the abelian $U(1)$ gauge invariance to
non-Abelian $SU(2)$-gauge invariance using the approximate conservation of
the isotopic spin current as starting point.\newline
In any case it is interesting to observe that the gauge principle originated
from General Relativity.\newline
Nowadays the notion of gauge symmetry is one of the cornerstones of
theoretical physics; the three non-gravitational interactions are described
by means of gauge theories in the framework of the Standard Model.\newline
Thanks to the works of Utiyama, Sciama and Kibble [25,27,28] it was realized
that also gravitation can be formulated as a gauge theory, the gauge group
being in that case the {\em Poincare` group,} that is the semidirect product
of the translation and the Lorentz group.\newline
More recently a quite general gauge theory has been formulated which
includes General Relativity as a particular case; in this case the gauge
group is the so called {\em affine group} resulting from the semidirect
product of the translation and the general linear group $GL(n,R)$. This
theory is called {\em Metric Affine Gravity} [22], and it allows for the
introduction of a general non-Riemannian connection.\newline
Though it is possible to treat gravity using the gauge approach, it is
necessary to remember that in the case of gravitation, contrary to the case
of strong and electroweak interactions, we are considering an {\it external}
symmetry group, i.e. a group acting on spacetime. So a procedure is
necessary to mediate the transition from the internal structure, which is
proper of any gauge formulation, and the external structures, and to project
geometric gauge structures on the base manifold in order to induce gravity.
It is presently unclear how this procedure applied to any affine frame takes
place. This issue is somewhat similar to the compactification of
higher-dimensional supergravity or string theories.\newline
Recently a different approach to metric affine gravity has been proposed by
Tucker and Wang [29-31] based on the metric ${\bf g}$ and the connection $%
{\bf \nabla }$ as independent variables. Instead of working with the group $%
GL(4,{\bf R)}$ (the general linear group), it relies on the definition of
torsion and non-metricity in terms of ${\bf g}$ and ${\bf \nabla }$.\newline

The study of non-Riemannian theories of gravitation is in general quite
complicate and a powerful formalism is needed in order to simplify
calculations. To this purpose, the frame-independent approach to
differential geometry seems quite appropriate. This approach has several
advantages. Indeed, we do not need to consider particular reference systems
or co-frames and the results are unambiguous and easy to apply, bearing in
mind that whenever requested, transition to the more traditional component
manipulation is possible ( taking into account the fact that since the
theory is non-metric, raising and lowering of indices must be done with
care).

\qquad Aim of this paper is to give an introduction to Metric Affine Gravity
in the Tucker-Wang approach. In particular, we shall review some particular
solutions of the Cartan equation for the non-riemannian part of the
connection. An application will be also given by deriving a Proca-type
equation for the trace of the non-metricity 1-forms. We shall consider only
the Cartan sector of the theory. This equivalence has been proved for the
Einstein sector too [34].\newline

The content of the paper is as follows. Some basic concepts and tools of non
riemannian geometry in the frame-independent formalism are reviewed in sect.
2. In sect. 3 we describe the variational techniques using a tensorial
approach. In sect. 4 some particular solutions of the Cartan equation are
obtained and classified. This classification will be necessary to prove (in
sect. 5) that a quite general model of non-Riemannian gravity yelds a Proca-
type equation for the Weyl 1-form $Q$.\newline
\qquad

\section{NON-RIEMANNIAN GEOMETRY}

\bigskip In this section a brief introduction is given of the non-Riemannian
geometry in the frame independent approach.\newline
The formalism we will be using takes into account the fact that in general
we will consider non-metric theories, i.e. theories in which the metric is
not regarded any longer as covariantly constant. It is clear that we need to
use a formalism which makes use of as less number of indices as
possible.This is obtained by formulating the non-Riemannian geometry in the
frame independent approach.\newline
One of the fundamental concepts in differential geometry is of Parallel
Transport. To define parallel transport we need to introduce a linear
connection, a type preserving derivation on the algebra of tensors fields
commuting with contractions. We will denote such a connection by ${\bf {%
\nabla }}$ . We can specify the most general linear connection by
calculating its effects on an arbitrary local basis of vector fields ${X_{a}}
$ 
\begin{equation}
{\nabla }_{X_{a}}X_{b}={\Lambda }^{c}{}_{b}(X_{a})X_{c}
\end{equation}
where $\Lambda ^{a}{}_{b}$ are a set of $n^{2}$ 1-forms, and $n$ is the
dimension of the manifold.\newline
It is possible to specify a general connection by giving a (2,0) metric
symmetric tensor ${\bf g}$, a (2,1) tensor ${\bf T}$ defined by 
\begin{equation}
{\bf T}(X,Y)={\bf \nabla }_{X}Y-{\bf \nabla }_{Y}X-[X,Y]
\end{equation}
(with $X,Y$ vector fields) and a (3,0) tensor ${\bf S}$\ symmetric in the
last two arguments . ${\bf T}$ is the torsion associated with ${\bf \nabla }$
and ${\bf S}$ is taken to be the metric gradient, ${\bf S=\nabla g}$. Then
it is possible to calculate the connection as a function of ${\bf g}$, ${\bf %
S}$, ${\bf T}$. Indeed, by using the relation 
\begin{equation}
X({\bf g}(Y,Z))={\bf S}(X,Y,Z)+{\bf g}({\bf {\nabla }}_{X}Y,Z)+{\bf g}(Y,%
{\bf \nabla }_{X}Z)
\end{equation}
we obtain 
\begin{eqnarray}
2{\bf g}(Z,{\bf \nabla }_{X}Y) &=&X({\bf g}(Y,Z))+Y({\bf g}(Z,X))-Z({\bf g}%
(X,Y))-{\bf g}(X,[Y,Z]) \\
&&-{\bf g}(Y,[X,Z])-{\bf g}(Z,[Y,X])-{\bf g}(X,{\bf T}(Y,Z))-{\bf g}(Y,{\bf T%
}(X,Z))-  \nonumber \\
&&{\bf g}(Z,{\bf T}(Y,X))-{\bf S}(X,Y,Z)-{\bf S}(Y,Z,X)+{\bf S}(Z,X,Y) 
\nonumber
\end{eqnarray}
where $X,Y,Z$ are any vector fields.\newline
We define the general curvature operator as: 
\begin{equation}
{\bf R}_{X,Y}Z={\bf \nabla }_{X}{\bf \nabla }_{Y}Z-{\bf \nabla }_{Y}{\bf %
\nabla }_{X}Z-{\bf \nabla }_{[X,Y]}Z
\end{equation}
which is a type preserving tensor derivation on the algebra of tensor
fields. The (3,1) curvature tensor is defined by: 
\begin{equation}
{\bf R}(X,Y,Z,\beta )=\beta (R_{X,Y}Z)
\end{equation}
with $\beta $ an arbitrary 1-form. We can introduce the following set of
local curvature 2-forms $R^{a}{}_{b}$: 
\begin{equation}
R^{a}{}_{b}(X,Y)=\frac{1}{2}{\bf R}(X,Y,X_{b},e^{a})
\end{equation}
where $e^{a}$ is any local basis of 1-forms dual to $X_{c}$. We have $%
e^{a}(X_{b})=\delta ^{a}{}_{b}$ or by using the contraction operator with
respect to $X$, $i_{X_{b}}(e^{a})=e^{a}(X_{b})=\delta ^{a}{}_{b}$.\newline
In terms of the connections forms we can write:\newline
\begin{equation}
R^{a}{}_{b}=d{\Lambda }^{a}{}_{b}+{\Lambda }^{a}{}_{c}\wedge {\Lambda }%
^{c}{}_{b}
\end{equation}
In a similar manner, the torsion tensor gives rise to a set of local 2-forms 
$T^{a}$ 
\begin{equation}
T^{a}(X,Y)\equiv \frac{1}{2}(e^{a}(T(X,Y))
\end{equation}
which can be written: 
\begin{equation}
T^{a}=de^{a}+\Lambda ^{a}{}_{b}\wedge e^{b}
\end{equation}
By using the symmetry of the tensor ${\bf g}$, the tensor ${\bf S}$ can be
used to define the local non-metricity 1-forms $Q_{ab}$ symmetric in their
indices: 
\begin{equation}
Q_{ab}(Z)={\bf S}(Z,X_{a},X_{b})
\end{equation}
It is convenient very often to make use of the exterior covariant derivative 
$D$.\newline
With $g_{ab}\equiv {\bf g}(X_{a},X_{b})$ we get that:\newline
\begin{eqnarray}
Q_{ab} &=&Dg_{ab} \\
Q^{ab} &=&-Dg^{ab}  \nonumber
\end{eqnarray}
As usual, indices are raised and lowered by means of the components of the
metric in a certain local basis. We denote the metric trace of these forms
as: 
\begin{equation}
Q=Q^{a}{}_{a}
\end{equation}
We call $Q$ the {\it Weyl 1-form}.\newline
In Riemannian geometry we require that the connection be metric compatible $%
(Q_{ab}=0$, or equivalently ${\bf S}=0)$ and ${\bf T}=0$.\newline
It is possible to decompose the connection ${\bf \nabla }$ into parts that
depend on the Levi-Civita connection ${\bf \stackrel{o}{\nabla }}$. To this
aim we introduce the tensor ${\bf \lambda }$ 
\begin{equation}
{\bf \lambda }(X,Y,\beta )=\beta ({\bf \nabla }_{X}Y)-\beta ({\bf \stackrel{o%
}{\nabla }}_{X}Y)
\end{equation}
for arbitrary vector fields $X,Y$ and 1-form $\beta $.\newline
To the decomposition above there corresponds a splitting of the connection
1-form into its Riemannian and non-Riemannian parts ${\Omega }^{a}{}_{b}$
and ${\lambda }^{a}{}_{b}$ , respectively, as: 
\begin{equation}
\Lambda ^{a}{}_{b}=\Omega ^{a}{}_{b}+\lambda ^{a}{}_{b}
\end{equation}
where 
\begin{equation}
\lambda ^{a}{}_{b}\equiv {\bf \lambda }(-,X_{b},e^{a})
\end{equation}
In terms of these forms we find: 
\begin{eqnarray}
T^{a} &=&\lambda ^{a}{}_{c}\wedge e^{c} \\
Q_{ab} &=&-(\lambda _{ab}+\lambda _{ba})  \nonumber
\end{eqnarray}
Using relations (11) and (17) we get $Q_{ab}={\bf S}(-,X_{a},X_{b})=-(%
\lambda _{ab}+\lambda _{ba})$, and therefore: 
\begin{equation}
{\bf S}=0\leftrightarrow \lambda _{ab}=-\lambda _{ba}
\end{equation}
namely the metric compatibility requires the antisymmetry of ${\bf \lambda \ 
}$.\newline
We define the 1-form $T$ by: 
\begin{equation}
T=i_{a}T^{a}
\end{equation}
with $i_{a}\equiv i_{X_{a}}$. We have the following relation: 
\begin{equation}
e^{c}\wedge \star T_{c}=-\star T
\end{equation}
where use has been made of the property $\star (A\wedge e^{a})=i^{a}\star A$
, with $A$ a generic form and $\star $ denotes the Hodge operation
associated with the metric ${\bf g}$.\newline
Since a general connection is neither symmetric nor antisymmetric,
particular care has to be taken when writing the indices, because in general 
$\lambda ^{a}{}_{b}$ is different from $\lambda _{b}{}^{a}$.\newline
Observe that a Riemannian connection $\Omega ^{a}{}_{b}$ being torsion free
implies: 
\begin{equation}
de^{a}+\Omega ^{a}{}_{b}\wedge e^{b}=0
\end{equation}
It follows from relation (4) that: 
\begin{equation}
2\Omega
_{ab}=(g_{ac}i_{b}-g_{bc}i_{a}+e_{c}i_{a}i_{b})de^{c}+(i_{b}dg_{ac}-i_{a}dg_{bc})e^{c}+dg_{ab}
\end{equation}
and 
\begin{equation}
2\lambda
_{ab}=i_{a}T_{b}-i_{b}T_{a}-(i_{a}i_{b}T_{c}+i_{b}Q_{ac}-i_{a}Q_{bc})e^{c}-Q_{ab}
\end{equation}
The field equations of the Einstein theory are obtained as variational
equations deduced from the Einstein-Hilbert action, this being the integral
of the curvature scalar of the Levi-Civita connection with respect to the
volume form.\newline
The scalar curvature is obtained by contracting the Ricci-tensor, which is
the trace of the curvature tensor.\newline
In general we can define two types of tensors: 
\begin{equation}
{\bf Ric}(X,Y)=e^{a}({\bf R}_{X_{a}X}Y)
\end{equation}
and 
\begin{equation}
{\bf ric}(X,Y)=e^{a}({\bf R}_{XY}X_{a})
\end{equation}
We have: 
\begin{eqnarray}
{\bf Ric}_{cb} &=&R_{acb}{}^{a} \\
{\bf ric}_{cb} &=&R_{cba}{}^{a}  \nonumber
\end{eqnarray}
where: 
\begin{equation}
R_{acb}{}^{d}={\bf R}(X_{a},X_{c},X_{b},e^{d})
\end{equation}
The first one has no symmetry in general while ${\bf ric}$ is a 2-form which
can be shown to be: 
\begin{equation}
{\bf ric}=2R^{a}{}_{a}=-dQ
\end{equation}
Indeed from relations (15), (22), (23) we get: 
\begin{equation}
2\Lambda =2{\Lambda }^{a}{}_{a}=-Q
\end{equation}
whereas relations (6,7) and (25) imply 
\begin{equation}
{\bf ric}=2R^{a}{}_{a}=2d\Lambda =-dQ
\end{equation}
since ${\Lambda }^{a}{}_{c}\wedge {\Lambda }^{c}{}_{a}=-{\Lambda }%
^{c}{}_{a}\wedge {\Lambda }^{a}{}_{c}=0$.\newline
In the Riemannian case ${\bf ric}=0$ , and ${\bf Ric}(X,Y)$ goes into the
usual Ricci tensor which is symmetric in the two arguments (by virtue of the
fact that non-metricity is zero); moreover the Riemann curvature tensor
satisfies the antisymmetry property 
\begin{equation}
R_{abc}{}^{d}=-R_{abd}{}^{c}
\end{equation}
The symmetric part of ${\bf Ric}$ can be contracted with the metric tensor
to obtain a generalised curvature: 
\begin{equation}
R={\bf Ric}(X_{a},X_{b}){\bf g}(X^{a},X^{b})={\bf Ri}c(X_{a},X^{a})
\end{equation}
It is possible to obtain the expression for the general ${\bf Ric}(X,Y)$ as: 
\begin{equation}
{\bf Ric}(X_{a},X_{b})={\stackrel{o}{Ric}}(X_{a},X_{b})+i_{a}i_{c}({%
\stackrel{o}{D}}\lambda ^{c}{}_{b}+\lambda ^{c}{}_{d}\wedge \lambda
^{d}{}_{b})
\end{equation}
where $\stackrel{o}{D}$ is the covariant exterior derivative with respect to
the Levi-Civita connection $\Omega ^{a}{}_{b}$ and 
\begin{equation}
R={\stackrel{o}{R}}+i_{a}i_{c}({\stackrel{o}{D}}\lambda ^{ca}+\lambda
^{c}{}_{d}\wedge \lambda ^{da})
\end{equation}
If we define the Ricci 1-forms by 
\begin{equation}
P_{b}=i_{a}R^{a}{}_{b}
\end{equation}
one has 
\begin{equation}
P_{a}={\bf Ric}(X_{b},X_{a})e^{b}
\end{equation}
and 
\begin{equation}
R=i^{b}P_{b}=i^{b}i_{a}R^{a}{}_{b}
\end{equation}
so that the curvature scalar can be written in general as: 
\begin{equation}
R=2R^{d}{}_{c}\otimes e^{c}\otimes X_{d}
\end{equation}

\section{Variations of the Generalised Einstein-Hilbert Action}

\bigskip A  non-Riemannian geometry is specified when we give a metric ${\bf %
g}$ and a connection ${\bf \nabla }$.\newline
In a local coframe ${e^{a}}$ with dual frame $X_{b}$ such that $%
e^{a}(X_{b})=\delta ^{a}{}_{b}$, the connection forms satisfy ($\Lambda
^{a}{}_{b}\equiv \omega ^{a}{}_{b}$): 
\begin{equation}
\omega ^{c}{}_{b}(X_{a})\equiv e^{c}(\nabla _{X_{a}}X_{b})
\end{equation}
In the following we use orthonormal frames so that: 
\begin{equation}
{\bf g}=\eta _{ab}e^{a}\otimes e^{b}
\end{equation}
with $\eta _{ab}=diag(-1,1,1,1,....)$.\newline
It is important to observe that the position (40) permits to transfer the
functional dependence on the metric ${\bf g}$ to the coframe $e^{a}$. Since
in general in the metric affine gauge theory of gravity [22] the metric $%
g_{ab}$ is considered as a gauge potential independent on $e^{a}$, the
assumption that $g_{ab}=\eta _{ab}$ is equivalent to choose a certain gauge
as well as requiring that ${\bf g}$ depends only on $e^{a}$. We will call it
the {\it Tucker-Wang gauge} [30,31].\newline
Let us consider an action written in the form: 
\begin{equation}
S[{\bf e},{\bf \omega }]=\int \Lambda ({\bf e},{\bf \omega })
\end{equation}
for some $n$-form $\Lambda $.\newline
The field equations of the theory follow from (mod d): 
\begin{eqnarray}
\underbrace{{\bf \Lambda }}_{{\bf e}} &=&0 \\
\underbrace{{\bf \Lambda }}_{{\bf \omega }} &=&0  \nonumber \\
&&  \nonumber
\end{eqnarray}
The general curvature scalar is: 
\begin{equation}
R=i^{b}i_{a}R^{a}{}_{b}
\end{equation}
The generalised Einstein-Hilbert action density $\Lambda _{EH}=R\star 1$ can
be written 
\begin{equation}
\Lambda _{EH}\equiv R\star 1=(i^{b}i_{a}R^{a}{}_{b})\star
1=R^{a}{}_{b}\wedge \star (e_{a}\wedge e^{b})
\end{equation}
>From the definition of curvature 2-forms it follows that: 
\begin{equation}
\underbrace{R\star 1}_{\omega }=(\underbrace{d\omega ^{a}{}_{b}+\omega
^{a}{}_{c}\wedge \omega ^{c}{}_{b}}_{\omega })\wedge \star (e_{a}\wedge
e^{b})
\end{equation}
We can write: 
\begin{equation}
d(\omega ^{a}{}_{b}\wedge \star (e_{a}\wedge e^{b}))=d\omega
^{a}{}_{b}\wedge \star (e_{a}\wedge e^{b})-\omega ^{a}{}_{b}\wedge d(\star
(e_{a}\wedge e^{b}))
\end{equation}
so that: 
\begin{eqnarray}
\underbrace{\Lambda _{EH}}_{\omega } &=&\dot{\omega}^{a}{}_{b}\wedge d\star
(e_{a}\wedge e^{b})+(\underbrace{\omega ^{a}{}_{c}\wedge \omega ^{c}{}_{b}}%
_{\omega })\wedge \star (e_{a}\wedge e^{b}) \\
+d(\dot{\omega}^{a}{}_{b}\wedge \star (e_{a}\wedge e^{b})) &=&(\dot{\omega}%
^{a}{}_{c}\wedge \omega ^{c}{}_{b}-\dot{\omega}^{c}{}_{b}\wedge \omega
^{a}{}_{c})\wedge \star (e_{a}\wedge e^{b})  \nonumber \\
+{\dot{\omega}}^{a}{}_{b}\wedge d\star (e_{a}\wedge e^{b})+d(\dot{\omega}%
^{a}{}_{b}\wedge \star (e_{a}\wedge e^{b})) &=&\dot{\omega}^{a}{}_{b}\wedge
\omega ^{b}{}_{c}\wedge \star (e_{a}\wedge e^{c})  \nonumber \\
&&-\dot{\omega}^{a}{}_{b}\wedge \omega ^{c}{}_{a}\wedge \star (e_{c}\wedge
e^{b})+{\dot{\omega}}^{a}{}_{b}\wedge d\star (e_{a}\wedge e^{b})+d(\dot{%
\omega}^{a}{}_{b}\wedge \star (e_{a}\wedge e^{b}))  \nonumber \\
&=&\dot{\omega}^{a}{}_{b}\wedge D\star (e_{a}\wedge e^{b})+d(\dot{\omega}%
^{a}{}_{b}\wedge \star (e_{a}\wedge e^{b}))  \nonumber
\end{eqnarray}
where $D$ is the exterior covariant derivative and ${\dot{\omega ^{a}{}_{b}}}
$ denotes the variation of $\omega ^{a}{}_{b}$ . Since $\dot{\omega}%
^{a}{}_{b}$ has compact support 
\begin{equation}
\int \underbrace{{\Lambda }_{EH}}_{\omega }=\int {\dot{\omega}^{a}{}_{b}}%
\wedge D\star (e_{a}\wedge e^{b})
\end{equation}
\bigskip The coframe variation gives: 
\begin{eqnarray}
\underbrace{{\Lambda }_{EH}}_{e} &=&\underbrace{R^{a}{}_{b}\wedge \star
(e_{a}\wedge e^{b})}_{e}= \\
R^{a}{}_{b}\wedge \underbrace{\star (e_{a}\wedge e^{b})}_{e} &=&\delta
e^{c}\wedge R^{a}{}_{b}\star (e_{a}\wedge e^{b}\wedge e_{c})  \nonumber
\end{eqnarray}
because $R^{a}{}_{b}$ is a coframe-independent object.\newline
We can write: 
\begin{equation}
\underbrace{\Lambda _{EH}}_{e}=\delta e^{c}\wedge G_{c}
\end{equation}
where the Einstein $(n-1)$-forms are: 
\begin{equation}
G_{c}=R^{a}{}_{b}\wedge \star (e_{a}\wedge e^{b}\wedge e_{c})
\end{equation}
For any coframe-independent $p$-forms $\alpha $ and $\beta $, the following
relation holds: 
\begin{equation}
\underbrace{\alpha \wedge \star \beta }_{e}=-\dot{e^{c}}\wedge \lbrack
i_{c}\beta \wedge \star \alpha -(-1)^{p}\alpha \wedge i_{c}\star \beta ]
\end{equation}
This can be proved as follows. Let us write the generic $p$-form $\beta $
as: 
\begin{equation}
\beta =\beta _{a_{1},a_{2},.....a_{p}}e^{a_{1}a_{2}....a_{p}}
\end{equation}
>From the frame-independence of $\beta $ it follows: 
\begin{equation}
\frac{1}{p}\delta _{e}({\beta _{a_{1},a_{2},.....a_{p}}}%
)e^{a_{1}a_{2}....a_{p}}+\beta _{a_{1},a_{2},.....a_{p}}(\delta
_{e}e^{a_{1}}\wedge e^{a_{2}....a_{p}})=0
\end{equation}
Then 
\begin{equation}
\delta _{e}(\star \beta )=\delta _{e}(\beta _{a_{1},a_{2},.....a_{p}}\star
e^{a_{1}a_{2}....a_{p}})
\end{equation}
so that 
\begin{eqnarray}
\underbrace{\alpha \wedge \star \beta }_{e} &=&\alpha \wedge \delta (\beta
_{a_{1},a_{2},.....a_{p}}\star e^{a_{1}a_{2}....a_{p}}) \\
&=&\alpha \wedge \delta _{e}(\beta _{a_{1},a_{2},.....a_{p}})\star
e^{a_{1}a_{2}....a_{p}}+\alpha \wedge \beta _{a_{1},a_{2},.....a_{p}}\delta
(\star e^{a_{1}a_{2}....a_{p}})  \nonumber \\
&=&\delta _{e}(\beta _{a_{1},a_{2},.....a_{p}})e^{a_{1}a_{2}....a_{p}}\wedge
\star \alpha +(-1)^{p}\delta e^{c}\wedge \alpha \wedge \beta
_{a_{1},a_{2},.....a_{p}}\star (e^{a_{1}a_{2}....a_{p}}{}_{c})  \nonumber \\
&=&-p\beta _{a_{1},a_{2},.....a_{p}}\delta e^{a_{1}}\wedge
e^{a_{2}....a_{p}}\wedge \star \alpha +(-1)^{p}\delta e^{c}\wedge \alpha
\wedge i_{c}\star \beta   \nonumber \\
&=&-\delta e^{c}\wedge \lbrack i_{c}\beta \wedge \star \alpha
-(-1)^{p}\alpha \wedge i_{c}\star \beta ]  \nonumber
\end{eqnarray}
where use has been made of the property $\alpha \wedge \star \beta =\beta
\wedge \star \alpha $ in the third line and of eq. (54) in the fourth line.%
\newline
>From eq. (52) we can get a relation between the scalar curvature and the
stress forms. To get it, notice that if $\alpha $ and $\beta $ are
frame-dependent then we have to add to (52) the coframe variation of $\alpha 
$ and $\beta $. Let us define $\Delta \tau _{c}$ by: 
\begin{equation}
\delta e^{c}\wedge \Delta \tau _{c}=\underbrace{\alpha }_{e}\wedge \star
\beta +\alpha \wedge \star \underbrace{\beta }_{e}
\end{equation}
Consider an action of the form: 
\begin{equation}
\Lambda =kR\star 1+\sum_{k}b_{k}(\alpha _{k}\wedge \star \beta _{k})
\end{equation}
where $\alpha _{k}$ and $\beta _{k}$ are generic $p_{k}$-forms and $b_{k}$
are constants.\newline
The coframe variation of (58) yields the equations: 
\begin{equation}
kR^{a}{}_{b}\wedge i_{c}i^{b}i_{a}\star 1-\sum_{k}b_{k}[i_{c}\beta
_{k}\wedge \star \alpha _{k}-(-1)^{p_{k}}\alpha _{k}\wedge i_{c}\star \beta
_{k}]+\sum_{k}\Delta \tau _{c}[k]=0
\end{equation}
where $\Delta \tau _{c}[k]$ is the extra term in the stress forms coming
from the generic term $b_{k}(\alpha _{k}\wedge \star \beta _{k})$.\newline
By taking the wedge product of (21) with $e^{c}$ we get: 
\begin{equation}
(-1)^{n+1}(n-2)R\star 1+(-1)^{n}\sum_{k}[(2p_{k}-n)b_{k}(\alpha _{k}\wedge
\star \beta _{k})+\sum_{k}(\Delta \tau _{c}[k])\wedge e^{c}=0
\end{equation}
\bigskip Consider now a situation in which we have an action density of the
form: 
\begin{equation}
\Lambda _{EH}+F({\bf e},{\bf \omega })
\end{equation}
The connection variation gives the equation: 
\begin{equation}
D\star (e^{a}\wedge e_{b})=F^{a}{}_{b}
\end{equation}
with $F^{a}{}_{b}$ ($n-1$)-forms defined by: 
\begin{equation}
\underbrace{F}_{\omega }=\dot{\omega}^{b}{}_{a}\wedge F^{a}{}_{b}
\end{equation}
Equation (62) is called {\it the Cartan equation}.\newline
>From (63) it follows 
\begin{equation}
F^{a}{}_{a}=0
\end{equation}
We can define the set of $0$-forms $f^{ca}{}_{b}$ by 
\begin{equation}
F^{a}{}_{b}=f^{ca}{}_{b}\star e_{c}
\end{equation}
>From (64) we get $f^{ca}{}_{a}=0$.\newline
It is possible to obtain the general solution of the Cartan equation by
decomposing both the Weyl form and the torsion into a trace part and a
traceless part [32]. We have, respectively: 
\begin{equation}
Q_{ab}={\hat{Q}}_{ab}+\frac{1}{n}g_{ab}Q
\end{equation}
where ${\hat{Q}}^{a}{}_{a}=0$, and  analogously 
\begin{equation}
T^{a}={\hat{T}}^{a}+\frac{1}{n-1}e^{a}\wedge T
\end{equation}
with $T\equiv i_{a}T^{a}$ and $i_{a}{\hat{T}}^{a}=0$ .\newline
Eq. (62) becomes therefore decomposed as: 
\begin{equation}
i_{b}\hat{Q}_{a}{}^{c}-\delta ^{c}{}_{b}i_{d}\hat{Q}_{a}{}^{d}+(\delta
^{c}{}_{b}\delta ^{d}{}_{a}-\delta ^{c}{}_{a}\delta ^{d}{}_{b})(\frac{n-2}{2n%
}i_{d}Q-i_{d}i_{h}T^{h})-i_{b}i_{a}T^{c}+f^{c}{}_{ab}=0
\end{equation}
that is, 
\begin{eqnarray}
i_{a}\hat{Q}_{bc}-i_{a}i_{b}T_{c} &=&-\frac{1}{2n}g_{bc}i_{a}Q+\frac{1}{2n}%
g_{ac}i_{b}Q-f_{cba} \\
&-&\frac{1}{n(n-2)}g_{ac}f^{d}{}_{db}+\frac{n-1}{n(n-2)}g_{bc}f^{d}{}_{da} 
\nonumber \\
&+&\frac{n-1}{n(n-2)}g_{ac}f^{d}{}_{bd}-\frac{1}{n(n-2)}g_{bc}f^{d}{}_{ad} 
\nonumber
\end{eqnarray}
Using the symmetry of ${\hat{Q}}_{ab}$ and the antisymmetry of $%
i_{a}i_{b}T_{c}$ , we find 
\begin{equation}
i_{a}i_{b}{\hat{T}}_{c}=\frac{1}{n-1}(g_{bc}f^{d}{}_{ad}-g_{ac}f^{d}{}_{bd})-%
\frac{1}{2}(f_{bac}+f_{bca}+f_{cab}-f_{cba}-f_{abc}-f_{acb})
\end{equation}
\begin{equation}
i_{a}{\hat{Q}}_{bc}=\frac{1}{n}g_{bc}(f^{d}{}_{da}+f^{d}{}_{ad})-\frac{1}{2}%
(f_{bac}+f_{bca}+f_{cab}+f_{cba}-f_{abc}-f_{acb})
\end{equation}
and 
\begin{equation}
T-\frac{n-1}{2n}Q=\frac{1}{n(n-2)}(f^{c}{}_{ac}+(1-n)f^{c}{}_{ca})e^{a}
\end{equation}
The following expressions hold true for the traceless parts of the torsion
and the nonmetricity, respectively: 
\begin{equation}
\hat{T_{c}}=\frac{1}{n-1}(e_{c}\wedge e^{a})f^{d}{}_{ad}-\frac{1}{2}%
(e^{b}\wedge e^{a})(f_{bac}+f_{bca}+f_{cab})
\end{equation}
\begin{equation}
\hat{Q_{ab}}=\frac{1}{n}%
g_{ab}(f^{d}{}_{da}+f^{a}{}_{ad})e^{a}-(f_{bac}+f_{bca}-f_{abc})e^{a}
\end{equation}
Equations (72)-(74) provide the general solution of the Cartan equation
(62). The study of the properties of such a general solution is fundamental
in the study of properties of non-Riemannian theories of gravitation [33].%
\newline
Let us notice that in general the connection variation of a generic action
gives a Cartan equation which is a differential equation to be solved for
the non-Riemannian part of the connection. In the present case, however, the
fact that we are considering an action density like (61) allows us to solve
algebraically eq. (62) for $\lambda ^{a}{}_{b}$.\newline
\newline

\section{Particular Solutions of the Cartan Equation}

\bigskip In this section we present some particular solutions of the Cartan
equation (62). The different cases are classified depending on the explicit
expression of the ($n-1$)-forms $F^{a}{}_{b}$.\newline
In particular, we will prove that, if $F^{a}{}_{b}$ takes the special form
of subsection 1.3.6 or 1.3.8, the traceless part of the torsion ${\hat{T}}%
^{a}$ turns out to be zero.\newline

\subsection{$F^{a}{}_{b}=0$}

The first case we are going to consider is when $F^{a}{}_{b}=0$.\newline
Then, using the equations of the previous section, we get: 
\begin{equation}
i_{a}{\hat{Q}}_{bc}=0
\end{equation}
that is ${\hat{Q}}_{bc}=0$ and 
\begin{equation}
i_{a}i_{b}{\hat{T}}_{c}=0
\end{equation}
which means ${\hat{T}}_{c}=0$. We can write: 
\begin{equation}
T^{a}=\frac{1}{n-1}(e^{a}\wedge T)
\end{equation}
\begin{equation}
T=\frac{n-1}{2n}Q
\end{equation}
The non-metricity 1-forms and the torsion 2-forms result to be: 
\begin{equation}
Q_{ab}=\frac{1}{n}g_{ab}Q
\end{equation}
\begin{equation}
T^{a}=\frac{1}{2n}(e^{a}\wedge Q)
\end{equation}
The non-Riemannian part of the connection takes the following very simple
form: 
\begin{equation}
{\lambda }_{ab}=-\frac{1}{2n}Qg_{ab}
\end{equation}
so the traceless part of the non-Riemannian part of the connection, defined
by 
\begin{equation}
{\hat{\lambda}}^{a}{}_{b}={\lambda }^{a}{}_{b}-\frac{1}{n}{\lambda }%
^{c}{}_{c}\delta ^{a}{}_{b}
\end{equation}
is zero. \newline
\newline

\subsection{$f_{cab}=-f_{cba}$}

\bigskip In this case the formulas of sect.3 yield 
\begin{equation}
{\hat{Q}}_{ab}=0
\end{equation}
\begin{equation}
T-\frac{n-1}{2n}Q=\frac{1}{n-2}f^{c}{}_{ac}e^{a}
\end{equation}
The traceless part of the torsion 2-forms is calculated to be \newline
\begin{equation}
{\hat{T_{c}}}=\frac{1}{n-1}(e_{c}\wedge e^{a})f^{d}{}_{ad}+\frac{1}{2}%
(e^{b}\wedge e^{a})f_{cba}
\end{equation}
thus providing the solution 
\begin{equation}
Q_{ab}=\frac{1}{n}g_{ab}Q
\end{equation}
\begin{equation}
T_{c}=\frac{1}{n-1}(e_{c}\wedge e^{a})f^{d}{}_{ad}+\frac{1}{2}(e^{b}\wedge
e^{a})f_{cba}+\frac{1}{n-1}e_{c}\wedge T
\end{equation}
\newline

\subsection{$F^{a}{}_{b}=\sum_{k}(e^{a}\wedge i_{b}\star A_{k})$}

\bigskip Let us now consider the case in which the forms $F^{a}{}_{b}$ can
be written as: 
\begin{equation}
F^{a}{}_{b}=\sum_{k}(e^{a}\wedge i_{b}\star A_{k})
\end{equation}
with $A_{k}$ set of 1-forms.\newline
One gets easily 
\begin{equation}
f_{cab}=[\sum_{k}(i_{c}(A_{k})g_{ab}-i_{a}(A_{k})g_{cb})]
\end{equation}
so that 
\begin{equation}
f_{cab}=-f_{acb}
\end{equation}
Therefore 
\begin{equation}
f^{c}{}_{ac}=[(1-n)i_{a}\sum_{k}A_{k}]
\end{equation}
and 
\begin{equation}
f^{c}{}_{ca}=0
\end{equation}
The condition $f^{ca}{}_{a}=0$ implies $(n-1)i^{c}\sum_{k}A_{k}=0$ i.e. $%
\sum_{k}A_{k}=0$.\newline
Then we have 
\begin{equation}
T=(n-1)(\frac{Q}{2n}-\frac{\sum_{k}(A_{k})}{n(n-2)})
\end{equation}
The calculation of $Q_{bc}$ gives: 
\begin{equation}
Q_{bc}=[\sum_{k}(-e_{b}i_{c}A_{k}-e_{c}i_{b}A_{k}+\frac{n+1}{n}g_{bc}A_{k})]+%
\frac{1}{n}g_{bc}Q
\end{equation}
Indeed, from the general expression (74), on account of the antisymmetry of $%
f_{cab}$, one finds 
\begin{equation}
Q_{bc}=\frac{1}{n}g_{bc}\,f^{d}{}_{ad}e^{a}+f_{abc}e^{a}+f_{acb}e^{a}+\frac{1%
}{n}Qg_{bc}
\end{equation}
and, from eq. (89): 
\begin{eqnarray}
f^{d}{}_{ad}e^{a} &=&(1-n)\sum_{k}A_{k} \\
f_{acb}e^{a} &=&[\sum_{k}(A_{k}g_{cb}-i_{c}(A_{k})e_{b})]  \nonumber \\
f_{abc}e^{a} &=&[\sum_{k}(A_{k}g_{cb}-i_{b}(A_{k})e_{c})]  \nonumber \\
&&  \nonumber
\end{eqnarray}
Using the previous relations we get the result (94) for $Q_{bc}$.
Analogously, from eq. (73) we find 
\begin{eqnarray}
\hat{T}_{c} &=&[\sum_{k}(\frac{1}{n-1}(e_{c}\wedge e^{a})(1-n)i_{a}(A_{k}) \\
&+&\frac{1}{2}(e^{b}\wedge
e^{a})[i_{a}A_{k}g_{bc}-i_{b}A_{k}g_{ac}+i_{c}A_{k}g_{bc}  \nonumber \\
&-&i_{b}A_{k}g_{ac}+i_{a}A_{k}g_{bc}-i_{c}A_{k}g_{ab}])  \nonumber \\
&=&\sum_{k}(-(e_{c}\wedge A_{k})+e^{b}\wedge (e^{a}\wedge
i_{a}A_{k})g_{bc}+(e^{a}\wedge
e^{b})i_{b}(A_{k})g_{ac})]=\sum_{k}(e_{c}\wedge A_{k})  \nonumber \\
&&  \nonumber
\end{eqnarray}
so the traceless part of the torsion comes to be: \newline
\begin{equation}
\hat{T_{c}}=[e_{c}\wedge \sum_{k}(A_{k})]
\end{equation}
\newline

\subsection{$F^{a}{}_{b}={\protect\delta }^{a}{}_{b}\star A$}

\bigskip With $A$ a generic 1-form in this case: 
\begin{equation}
f_{cab}=g_{ab}i_{c}A
\end{equation}
The condition $f^{ca}{}_{a}=0$ gives $n\,i^{c}A=0$.\newline
Notice that: 
\begin{equation}
T-\frac{n-1}{2n}Q=-\frac{A}{n}
\end{equation}
Using the same method of the previous case we get: 
\begin{equation}
\hat{T_{c}}=\frac{n}{n-1}(e_{c}\wedge A)
\end{equation}
so that: 
\begin{equation}
T_{c}=\frac{n+1}{n}(e_{c}\wedge A)+\frac{1}{2n}(e_{c}\wedge Q)
\end{equation}
The non-metricity can be calculated using the expression (74) which, on the
basis of the symmetry properties of $f_{cab}$ , becomes: 
\begin{equation}
i_{a}{\hat{Q}}_{bc}=\frac{2}{n}g_{bc}f^{d}{}_{da}-f_{bac}-f_{cab}+f_{acb}
\end{equation}
or 
\begin{equation}
{\hat{Q}}_{bc}=\frac{2}{n}%
g_{bc}f^{d}{}_{da}e^{a}-f_{bac}e^{a}-f_{cab}e^{a}+f_{acb}e^{a}
\end{equation}
By plugging in the expression for $f_{abc}$ we get: 
\begin{eqnarray}
{\hat{Q}_{bc}} &=&[\frac{2}{n}g_{bc}i_{a}(A)e^{a}-g_{ac}(i_{b}A)e^{a} \\
&-&g_{ab}(i_{c}A)e^{a}+g_{cb}(i_{a}A)e^{a}  \nonumber \\
&=&\frac{2}{n}g_{bc}A-i_{b}Ae_{c}-i_{c}Ae_{b}+g_{bc}A  \nonumber \\
&=&-e_{c}i_{b}A-e_{b}i_{c}A+\frac{2+n}{n}g_{bc}A  \nonumber \\
&&  \nonumber
\end{eqnarray}
and finally 
\begin{equation}
Q_{bc}=[-e_{b}i_{c}(A)-e_{c}i_{b}(A)+\frac{n+2}{n}g_{bc}A]+\frac{1}{n}Qg_{bc}
\end{equation}
The last two cases are useful to understand the relation between certain
models of non-Riemannian gravity and Einstein theory [33]. \newline

\subsection{$F^{a}{}_{b}=e^{a}\wedge \star A_{b},_{{}}i^{b}A_{b}=0$}

\bigskip Here $A_{b}$ is a 2-form, wich can be assumed to be traceless
without loss of generality. Indeed, a not traceless $A_{b}$ can be written
as: 
\begin{equation}
A_{b}=\hat{A_{b}}+\frac{1}{n-1}(e_{b}\wedge A)
\end{equation}
with $A=i^{a}A_{a}$. The contribution from the second term gives a term of
the type already treated in 4.3, so we can limit ourselves to considering
the traceless case.\newline
It is easy to find that: 
\begin{equation}
f_{cab}=i_{a}i_{c}A_{b}
\end{equation}
whence 
\begin{eqnarray}
f_{cab} &=&-f_{acb} \\
f^{ca}{}_{a} &=&i^{c}i^{a}(A_{a})=0  \nonumber \\
f^{cc}{}_{a} &=&0  \nonumber \\
f^{ca}{}_{c} &=&i^{a}i^{c}(A_{c})=0  \nonumber
\end{eqnarray}
We get therefore 
\begin{equation}
T-\frac{n-1}{2n}Q=0
\end{equation}
In this case the relation between $Q$ and $T$ is the same as (78).\newline
After some calculations we find: 
\begin{equation}
{\hat{Q}}_{ab}=-[i_{a}(A_{b})+i_{b}(A_{a})]
\end{equation}
and 
\begin{equation}
\hat{T_{c}}=-[e^{a}\wedge i_{c}(A_{a})+A_{c}]
\end{equation}
which means 
\begin{equation}
e^{c}\wedge \hat{T_{c}}=e^{a}\wedge A_{a}
\end{equation}
If we consider the case in which $A_{b}$ can be written as 
\begin{equation}
A_{b}=i_{b}B
\end{equation}
with $B$ a 3-form, then $\hat{Q}_{ab}=0.$\newline
Another interesting case is when $A_{b}$ coincides with $\hat{T}_{b}$ apart
from a constant factor $\lambda $: $A_{b}=\lambda \hat{T}_{b}$. Then, it
follows from relation (113) that, if $\lambda =1$, $\hat{T}_{c}$ is
arbitrary, while if $\lambda \neq 1$ we need: 
\begin{equation}
e^{c}\wedge \hat{T}_{c}=0
\end{equation}

\subsection{(case 3.3 + 3.4) $F^{a}{}_{b}={\protect\delta }^{a}{}_{b}\star
A+\sum_{k}(e^{a}\wedge i_{b}\star A_{k})$}

\bigskip We get: 
\begin{equation}
f_{cab}=\sum_{k}((i_{c}A_{k})g_{ab}-(i_{a}A_{k})g_{cb})+g_{ab}(i_{c}A)
\end{equation}
The condition $f^{ca}{}_{a}=0$ gives: 
\begin{equation}
(n-1)\sum_{k}A_{k}+nA=0
\end{equation}
Therefore 
\begin{equation}
T=\frac{n-1}{2n}Q-[\frac{n-1}{n(n-2)}\sum_{k}A_{k}+\frac{1}{n}A]
\end{equation}
The non-metricity reads: 
\begin{equation}
Q_{bc}=[-e_{b}i_{c}(\sum_{k}A_{k}+A)-e_{c}i_{b}(\sum_{k}A_{k}+A)+\frac{n+1}{n%
}g_{bc}\sum_{k}A_{k}+\frac{n+2}{n}g_{bc}A]+\frac{1}{n}Qg_{bc}
\end{equation}
or, by using relation (118): 
\begin{equation}
Q_{bc}=[-e_{b}i_{c}(\sum_{k}A_{k}+A)-e_{c}i_{b}(\sum_{k}A_{k}+A)+\frac{2}{n}%
(\sum_{k}A_{k}+A)g_{bc}]+\frac{1}{n}g_{bc}Q
\end{equation}
By putting $A_{1}=-(\sum_{k}A_{k}+A)$ we can write: 
\begin{equation}
Q_{bc}=e_{b}i_{c}A_{1}+e_{c}i_{b}A_{1}-\frac{2}{n}A_{1}g_{bc}+\frac{1}{n}%
g_{bc}Q
\end{equation}
The traceless part of the torsion is found from (73) and (118): 
\begin{equation}
\hat{T}_{c}=[e_{c}\wedge \sum_{k}A_{k}+\frac{n}{n-1}e_{c}\wedge A]=0
\end{equation}
whence 
\begin{equation}
T^{c}=\frac{1}{n-1}(e^{c}\wedge T)
\end{equation}
\newline

\subsection{$F^{a}{}_{b}=\sum_{k}(e_{b}\wedge i^{a}\star A_{k})$}

\bigskip In this case we get: 
\begin{equation}
f_{cab}=[\sum_{k}(i_{c}(A_{k})g_{ab})-i_{b}(A_{k})g_{ac}]
\end{equation}
and therefore 
\begin{equation}
f_{cab}=-f_{bac}
\end{equation}
We see that $f^{c}{}_{ac}=0$ and: 
\begin{equation}
f^{c}{}_{ca}=[(1-n)i_{a}\sum_{k}A_{k}]
\end{equation}
The condition $f^{ca}{}_{a}=0$ implies $(n-1)i^{c}\sum_{k}A_{k}=0$ , so that 
\begin{equation}
T-\frac{n-1}{2n}Q=\sum_{k}[\frac{(n-1)^{2}}{n(n-2)}A_{k}]
\end{equation}
The calculation of $Q_{bc}$ yields the same expression already found for the
case 4.3: 
\begin{equation}
Q_{bc}=[\sum_{k}(-e_{b}i_{c}A_{k}-e_{c}i_{b}A_{k}+\frac{n+1}{n}g_{bc}A_{k})]+%
\frac{1}{n}g_{bc}Q
\end{equation}
As far as the traceless part of the torsion is concerned , one finds 
\begin{equation}
\hat{T}_{c}=\frac{1}{2}(e^{b}\wedge e^{a})f_{bca}=-(e^{b}\wedge
e^{a})g_{ca}\sum_{k}i_{b}A_{k}=e^{c}\wedge \sum_{k}A_{k}
\end{equation}
i.e. the same expression of the case 4.3.\newline
\newline

\subsection{(case 3.7 + 3.4) $F^{a}{}_{b}=\protect\delta ^{a}{}_{b}\star
A+\sum_{k}(e_{b}\wedge i^{a}\star A_{k})$}

\bigskip We have: 
\begin{equation}
f_{cab}=[\sum_{k}(i_{c}(A_{k})g_{ab}-i_{b}(A_{k})g_{ac})+g_{ab}i_{c}A
\end{equation}
It is easy to see that all the results of  the case 4.6 still hold , apart
from the relation between $T$ and $Q$ which is modified into: 
\begin{equation}
T=\frac{n-1}{2n}Q+[\frac{(n-1)^{2}}{n(n-2)}A_{k}-\frac{A}{n}]
\end{equation}
The expression for $Q_{bc}$ and $T_{c}$ are the same, whereas the traceless
part of the torsion 2-forms comes out to be zero.

\section{Applications: Proca theories from Metric Affine gravity}

\bigskip In this section we want to show briefly how, by using the findings
of the previous section, a quite general model in non-Riemannian gravity
gives a Proca-type equation for the Weyl form $Q$ . Such a result is
fundamental in proving the Obukhov theorem [34,35], and therefore the
content of this section is more than a simple application of the formalism.%
\newline
Let us consider a model in which the action is: 
\begin{eqnarray}
S &=&\int kR\star 1+\frac{c_{1}}{2}(dQ\wedge \star dQ)+\frac{c_{2}}{2}%
(Q\wedge \star Q)+\frac{c_{3}}{2}(Q\wedge \star T) \\
&&+\frac{c_{4}}{2}(T\wedge \star T)+\frac{c_{5}}{2}(T^{c}\wedge \star T_{c})+%
\frac{c_{6}}{2}({\cal {Q}}\wedge \star Q)+  \nonumber \\
&&\frac{c_{7}}{2}({\cal {Q}}\wedge \star T)+\frac{c_{8}}{2}({\cal {Q}}\wedge
\star {\cal {Q})}  \nonumber
\end{eqnarray}
where: 
\begin{equation}
{\cal {Q}}={e^{a}}{i^{b}}{Q_{ab}}
\end{equation}
We want to show that, irrespective of the values of coefficients $%
c_{1},...c_{8}$ , one gets a Proca equation for $Q$.\newline
Let us derive the connection variation of all terms in the action (132). The
results one easily gets (mod d) are as follows:

\subparagraph{ Terms $dQ\wedge \star dQ,(Q\wedge \star Q),(T\wedge \star
T),(T^{c}\wedge \star T_{c})$:}

\begin{equation}
\delta \omega ^{a}{}_{b}\wedge \lbrack 2\delta ^{b}{}_{a}(c_{1}d\star
dQ+c_{2}\star Q)+c_{4}(e^{b}\wedge i_{a}\star T)-c_{5}(e^{b}\wedge \star
T_{a})]
\end{equation}

\subparagraph{Term $\frac{c_{3}}{2}(Q\wedge \star T)$:}

\begin{equation}
-2\delta \omega ^{a}{}_{b}\delta ^{a}{}_{b}\wedge \star T-\delta \omega
^{a}{}_{b}\wedge (e^{b}\wedge i_{a}\star Q)
\end{equation}

\bigskip 

\subparagraph{Term ${\cal {Q}\wedge \star T}$  :}

\begin{eqnarray}
&&-2\delta \omega ^{a}{}_{b}\delta ^{a}{}_{b}\wedge \star T-\delta \omega
^{a}{}_{b}\wedge (e^{b}\wedge i_{a}\star {\cal {Q})+} \\
&&\delta \omega ^{a}{}_{b}\wedge (e_{a}\wedge i^{b}\star T)+\delta \omega
^{a}{}_{b}\wedge (e^{b}\wedge i_{a}\star T)  \nonumber
\end{eqnarray}

\subparagraph{Term ${\cal {Q}}\wedge \star Q$ :}

\begin{equation}
-2\delta \omega ^{a}{}_{b}\delta ^{b}{}_{a}\wedge \star (Q+{\cal {Q}}%
)+\delta \omega ^{a}{}_{b}\wedge (e_{a}\wedge i^{b}\star Q)+\delta \omega
^{a}{}_{b}\wedge (e^{b}\wedge i_{a}\star Q)
\end{equation}

\subparagraph{Term ${\cal {Q}}\wedge \star {\cal {Q}}$ :}

\begin{equation}
-4\delta \omega ^{a}{}_{b}\delta ^{b}{}_{a}\wedge \star {\cal {Q}}+2\delta
\omega ^{a}{}_{b}\wedge (e_{a}\wedge i^{b}\star {\cal {Q}})+2\delta \omega
^{a}{}_{b}\wedge (e^{b}\wedge i_{a}\star {\cal {Q}})
\end{equation}
The equation we get from the connection variation will split in a trace part
and a traceless part. \newline
All the terms considered in (134)-(138) are of the type considered in
section 4. Then, it is easy to check that in general the trace part will
contain terms like 
\begin{equation}
\lambda _{1}\star T+\lambda _{2}\star Q+\lambda _{3}\star {\cal {Q}}
\end{equation}
where $\lambda _{1},\lambda _{2},\lambda _{3}$ are constants.\newline
Then we can write: 
\begin{equation}
c_{1}d\star dQ+c_{2}\star Q=\lambda _{1}\star T+\lambda _{2}\star Q+\lambda
_{3}\star {\cal {Q}}
\end{equation}
where $\lambda _{1},\lambda _{2},\lambda _{3}$ are constants related to $%
c_{3},c_{4},....c_{8}$.\newline
We can get other two independent relations from the  condition $%
f^{ca}{}_{a}=0$ and from (72): \bigskip 
\begin{equation}
T-\frac{n-1}{2n}Q=\frac{1}{n(n-2)}(f^{c}{}_{ac}+(1-n)f^{c}{}_{ca})e^{a}
\end{equation}
Exploiting the relations of the previous sections it is possible to verify
that these two conditions in general will give two independent linear
relations between $Q,T,{\cal {Q}}$ which can be used to eliminate $T,{\cal {Q%
}}$ in eq. (140).\newline
The conclusion is that Eq. (140) can be written as: 
\begin{equation}
c_{1}d\star dQ+(c_{2}+\Delta c_{2})\star Q=0
\end{equation}
where $\Delta c_{2}$ will in general depend on the constants $%
c_{3},c_{4},.....c_{8}$.\newline
Therefore, the Proca-type behaviour for $Q$ is proved. Also the fields $T,%
{\cal {Q}}$ will satisfy Proca-type equations.\newline
Let us observe that by choosing the constants $c_{3},c_{4},....c_{8}$ in a
proper way we can satisfy the condition: 
\begin{equation}
c_{2}+\Delta c_{2}=0
\end{equation}
so that we get a Maxwell-like equation for $Q$: 
\begin{equation}
d\star dQ=0
\end{equation}
Such a result can be used to get exact solutions in Metric Affine Gravity
from known solutions of Einstein-Maxwell theory [36,37].\newline

{\bf ACKNOWLEDGMENTS}\newline
\newline
One of us (R.S.) thanks the NOOPOLIS Foundation, Italy, for financial
support, and R. Tucker and C. Wang for stimulating discussions on the topic.%
\newline
\newpage 

\begin{center}
{\bf REFERENCES}
\end{center}

\bigskip 1] E. Kroener, {\em Continuum theory of defects}, in {\it Physics
of Defects, Les Houches Session XXXV}, 1980, R. Balian et al. eds.
(North-Holland, Amsterdam 1981), p. 215.\newline
\newline
2] E. Kroener, {\em The role of differential geometry in the mechanics of
solids}, in {\it Proc. 5th Nat. Congr. Theor. Appl. Mech.,} Vol. 1,
(Bulgarian Acad. Sci. Sofia 1985), p. 352 .\newline
\newline
3] Y. Neeman, D. Sijacki, Ann. Phys. {\bf 120} (1979) 292.\newline
\newline
4] Y. Neeman, D. Sijacki, Phys. Rev. D {\bf 37} (1988) 3267.\newline
\newline
5] P. J. Steinhart, Class. Quant. Grav. {\bf 10} (1993) S33.\newline
\newline
6] J. Scherk and J. H. Schwartz, Phys. Lett. B {\bf 52B} (1974) 347.\newline
\newline
7] T. Dereli, R. W. Tucker, Class. Quant. Grav. {\bf 12} L31 (1995).\newline
\newline
8] T. Dereli, M. Onder, R. W. Tucker, Class. Quant. Grav. {\bf 12} L25
(1995).\newline
\newline
9] T. Dereli, R. Tucker, Class. Quant. Grav. {\bf 11} (1994) 2575.\newline
\newline
10] F. W. Hehl, E. Lord, L. L. Smalley, Gen. Rel. Grav. {\bf 13} (1981) 1037.%
\newline
\newline
11] F. W. Hehl, E. W. Mielke: {\it Non-metricity and Torsion}, {\em Proc.
4th Marcel Grossman Meeting on General relativity}, Part A, ed. R. Ruffini (
North Holland, Amsterdam, 1986) p. 277.\newline
\newline
12] V. N. Ponomariev, Y. Obukhov, Gen. Rel. Grav. {\bf 14} (1982) 309.%
\newline
\newline
13] A. A. Coley, Phys. Rev. D, {\bf 27} (1983) 728.\newline
\newline
14] A. A. Coley, Phys. Rev. D. {\bf 28} (1983) 1829.\newline
\newline
15] A. A. Coley, Nuovo Cim. {\bf B69} (1982) 89.\newline
\newline
16] M. Gasperini, Class. Quant. Grav. {\bf 5} (1988) 521.\newline
\newline
17] J. Stelmach, Class. Quant. Grav. {\bf 8} (1991) 897.\newline
\newline
18] A. K. Aringazin, A. L. Mikhailov, Class. Quant. Grav. {\bf 8} (1991)
1685.\newline
\newline
19] J. P. Berthias. B. Shabid-Saless, Class. Quant. Grav. {\bf 10} (1993)
1039.\newline
\newline
20] L. L. Smalley, Phys. Rev. D {\bf 21} (1980) 328.\newline
\newline
21] R. Tucker, C. Wang, Class. Quant. Grav. {\bf 15} (1998) 933.\newline
\newline
22] F. W. Hehl, J. D. McCrea, E. W. Mielke, Y. Neeman. Phys. Rep. {\bf 258}
1 (1995).\newline
\newline
23] Yu. Obukhov, V. N. Ponomariev, V.V. Zhytnikov, Gen. Rel. Grav. {\bf 21}
(1989) 1107.\newline
\newline
24] J. F. Pascual-Sanchez, Phys. Lett. A {\bf 108} (1985) 387.\newline
\newline
25] D. W. Sciama, Rev. Mod. Phys. {\bf 36} (1964) 463 and 1103.\newline
\newline
26] H. Weyl, {\it Geometrie und Physik}, Naturwissenschaften, {\bf 19}
(1931) 49.\newline
\newline
27] R. Utiyama, Phys. Rev. {\bf 101} (1956) 1597.\newline
\newline
28] T. W. B. Kibble, J. Math. Phys. {\bf 2} (1961) 212.\newline
\newline
29] C. Wang, Ph. D. thesis (Lancaster) (1996).\newline
\newline
30] R. Tucker, C. Wang, {\em Non Riem. Grav. Interac.}, {\it Mathematics of
Gravitation}, Banach Centre Publications, Vol. 41 (Warsawa ,1997).\newline
\newline
31] R. W. Tucker, C. Wang, Class. Quant. Grav. {\bf 12} (1995) 2587.\newline
\newline
32] J. D. McCrea, Class. Quant. Grav. {\bf 9} (1992) 553.\newline
\newline
33] R. Scipioni, Ph. D. Thesis (Lancaster) (1999).\newline
\newline
34] Yu. N. Obukhov, E. J. Vlachynsky, W. Esser and F. W. Hehl, Phys. Rev. D 
{\bf 56} (1997) 7769.\newline
\newline
35] R. Scipioni, J. Math. Phys. {\bf 41} 5 (2000).\newline
\newline
36] T. Dereli, M. Onder. J. Schray, R. W. Tucker, C. Wang, Class. Quant.
Grav. {\bf 13} (1996) L103.\newline
\newline
37] F. W. Hehl, A. Macias, Int. J. Mod. Phys. D, {\bf 8} 4 (1999) 399.
\newline

\end{document}